\documentclass[12pt]{article}
\title{
On dynamics of fermion generations }
\author{Yu.A.Simonov\\ Institute of Theoretical and Experimental
Physics\\ 117118, Moscow, B.Cheremushkinskaya 25, Russia}
\date{}

\def\la{\mathrel{\mathpalette\fun <}}
\def\ga{\mathrel{\mathpalette\fun >}}
\def\fun#1#2{\lower3.6pt\vbox{\baselineskip0pt\lineskip.9pt
\ialign{$\mathsurround=0pt#1\hfil
##\hfil$\crcr#2\crcr\sim\crcr}}}

\newcommand{\beq}{\begin{eqnarray}}
 \newcommand{\eeq}{\end{eqnarray}}
\newcommand{\be}{\begin{equation}}
 \newcommand{\ee}{\end{equation}}

 \def\la{\mathrel{\mathpalette\fun <}}
\def\ga{\mathrel{\mathpalette\fun >}}
\def\fun#1#2{\lower3.6pt\vbox{\baselineskip0pt\lineskip.9pt
\ialign{$\mathsurround=0pt#1\hfil ##\hfil$\crcr#2\crcr\sim\crcr}}}

\newcommand{{\SD}}{\rm SD}

\newcommand{\lan}{\langle}
\newcommand{\ran}{\rangle}

\newcommand{\wbp}{\widehat{\bar\psi}}
\newcommand{\wps}{\widehat{\psi}}

\begin{document}
\maketitle
\begin{abstract}
The hierarchy of fermion masses and EW symmetry breaking without elementary
Higgs is studied on the basis of strong gauge field distributions governing the
EW dynamics. The mechanism of symmetry breaking due to quark bilinears
condensation is generalized to the case, when higher field  correlators are
present in the EW vacuum. Resulting wave functional yields several minima of
quark bilinears, giving masses of three (or more) generations.  Mixing is
suggested to be  due to   kink solutions of the same wave functional. For a
special form of this mixing ("coherent mixing") a realistic hierarhy of masses
and CKM coefficients is obtained and arguments in favor of the fourth
generation are given. Possible important role of topological charges for CP
violating phases and small masses of the first generation is stressed.
\end{abstract}

\section{Introduction}

Despite spectacular success of the standard model (SM), the Higgs
sector and the pattern of fermion masses and mixing remains mostly
an unsolved issue, for the theoretical overview   see [1]. In this
paper we suggest a framework which might shed some light on the
origin of generations, the hierarchy of fermion masses, and the
Higgs problem.

The topics mentioned above are related to several related
problems:

i) Dynamical origin of Higgs sector and spontaneous symmetry
breaking in  the $SU(2)\times U(1)$ sector;

ii) Fermion generations and  hierarchy of fermion mass matrices;

iv) Origin of CP violation.

Possible solutions of the Higgs problem, different from the
popular SUSY scenario, have been suggested by technicolor model
[2] and by economical idea of top condensate [3,4,5], with a
modern development of topcolor--assisted technicolor model [6].
There a strong interaction at high scale $M \sim 10^{15} -
10^{16}$ GeV allows to create Higgs sector dynamically, but leaves
points ii) and iv) unsolved.

A way to understand large top mass  was suggested already 20 years
ago [7] and developed in detailed manner since then [8,9]. The
symmetry responsible for large top mass was called "flavour
democracy" and  considered in family space in each of the sectors
(up, down and leptons) separately. The realization of this
symmetry in the framework of the "flavor gauge theory" was given
in [10].

The flavor--democratic scenario  illustrates why in each of the
sectors the mass of the  third family is much larger than that in
first two families and allows to connect phenomenologically the
CKM mixing angles with masses [11]. However, it does not consider
dynamical origin of first two families and another important
hierarchy: why scales of the masses in three sectors are so much
different, and inside the family the mass of top is much larger,
than that of bottom and tau--lepton.

Summarizing, the problem of lower generations was not addressed.
It is remarkable that masses of the first generation have much
smaller scale, which might signify that internal dynamics may
differ from generation to generation. Also the dynamical mechanism
producing generations remains unknown. It is a purpose of present
paper to suggest a possible variant of such mechanism, based on
nonperturbative dynamics  of the EW gauge  and fermion fields. We
will show below that the fermion masses due to Spontaneous
Symmetry Breaking (SSB) naturally form the family structure, when
higher field correlators are taken into account.

We also show that dilute topological charges in the EW vacuum may
be responsible for the dynamics of the lowest fermion family. This
formalism can be used  for producing Higgs phenomenon in the same
way as it was done in the topcolor-type models \cite{3,4,5,6}.

In this way the Higgs is coupled to (and made of ) all fermions,
and the scalar condensate is formed dynamically, giving mass to
all quarks. The field--theoretical framework allows to consider
additional contributions from topological charges  creating
nonzero masses for light fermions of first generation. The fermion
mixing is associated with the kink solutions of the same wave
functional, which connect different stationary points
corresponding to generations. For a special form of the mass
matrix, called the coherent mixing form, the mass eigenvalues have
a pronounced hierarchy and CKM mixing coefficients are expressed
via the mass ratios yielding realistic values. The neutrino mass
can be considered on the same ground, including leptons and quarks
symmetrically, and then  the mixing, both in quark and lepton
sectors, is obtained in the same way.

The plan of the  paper  is as follows. In Section 2 the gauge interaction at
intermediate scale is introduced and resulting multifermion Lagrangian is
derived. The gap equation is solved and the  mass matrix is obtained and
discussed  in Section 3. Contributions of topological charges are given in
Section 4. The problem of the fermion mixing is studied  in Section 5, while
the SSB and Higgs dynamics is presented is Section 6. Section 7  is devoted to
a summary and possible developments of the method. Three Appendices contain an
additional material for derivation of formulas in the text: Appendix 1 yields
the quark Green's function in the field of topological charges; Appendices 2
and 3 describe diagonalization of the mass matrices in the case of three and
four generations.

\section{Derivation of the multi--fermion Lagrangian}

The SM Lagrangian can be split in two parts,
 \be L_{SM} = L_{st}+ L_{Higgs}, \ee
where $L_{st}$ contains all kinetic parts  of fermions and gauge bosons and
their interaction, whereas $L_{Higgs}$ refers to all terms where the Higgs
field appears. It is our purpose, as in Refs.~\cite{3}--\cite{5}, to derive
$L_{Higgs}$   with effective Higgs field from the fields present in  $L_{st}$,
which would generate dynamically Higgs condensate, fermion masses, and mixings.

To this end, first of all,  we  must organize fermions into some
structures which enter the fundamental Lagrangian, namely,
$$
\psi\epsilon\{ \psi^A_{Li}, \psi^A_R\},~~ A=\{n,\alpha\},
$$
where $\alpha=1,2,3$ refers to families and $n=1,2,3,4$ refers to
" sectors" of fermions, which can be composed as follows \be
\psi^{1,\alpha}_{Li}= \left (
\begin{array}{l}
e^c_L\\
\nu^c_{eL}
\end{array}
\right ) ,
\left (
\begin{array}{l}
\mu^c_L\\
\nu^c_{\mu L}
\end{array}
\right ) ,
\left (
\begin{array}{l}
\tau^c_L\\
\nu^c_{\tau L}
\end{array}
\right ) , \psi^{1,\alpha}_R= (\nu^c_{eR},\nu^c_{\mu R},
\nu^c_{\tau R})
 \ee \be \psi^{2,\alpha}_{Li}= \left (
\begin{array}{l}
\nu_{eL}\\
e_L
\end{array}
\right ) ,
\left (
\begin{array}{l}
\nu_{\mu L}\\
\mu_ L
\end{array}
\right ) ,
\left (
\begin{array}{l}
\nu_{\tau L}\\
\tau_L
\end{array}
\right ) , \psi^{2,\alpha}_R= (e_R,\mu_R,\tau_R)
 \ee \be
\psi^{3,\alpha}_{Li}= \left (
\begin{array}{l}
u_{L}\\
d_L
\end{array}
\right ) ,
\left (
\begin{array}{l}
c_L\\
s_L
\end{array}
\right ) ,
\left (
\begin{array}{l}
t_L\\
b_L
\end{array}
\right ) ,
\psi^{3,\alpha}_R=
(d_R,s_R,b_R)
\ee
\be
\psi^{4,\alpha}_{Li}=
\left (
\begin{array}{l}
d^c_L\\
u^c_L
\end{array}
\right ) ,
\left (
\begin{array}{l}
s^c_L\\
c^c_L
\end{array}
\right ) ,
\left (
\begin{array}{l}
b^c_L\\
t^c_L
\end{array}
\right ) ,
\psi^{4,\alpha}_R=
(u^c_R,c^c_R,t^c_R)
\ee

Note that considering  the gauge dynamics at high scale $M$, one
can introduce, similarly to [11], the "urfermions" with quantum
numbers which are possibly different from those of final
diagonalized fermions in (2-5).

Urfermions are denoted by the hat sign, $\widehat\psi^A_a$ and
supplied by an additional index $a$, implying that $\widehat \psi$
belongs to a representation of some gauge group $G$ operating at
the scale $M$. We shall assume here that this group is broken at
low scale and only one, the lowest mass component $a=1$, should be
considered  for low  scale dynamics. The diagonalized form of
$\widehat \psi^A_{a=1}$ will be associated with the physical
states listed in (2-5).

The destiny of higher states, $\widehat \psi_a,  a=2,3..$ will be
discussed elsewhere, together with a possibility that the sets
$\{a\}$ and $\{A\}$ have a common intersection. In what follows we
consider the  simplest case with  standard fermions listed in
sectors (2-5).

The fundamental Lagrangian at high scale reads ( the Euclidean fields and
metrics are used everywhere) \be L_{high}=g\widehat{\bar{\psi}}^A_a
(x)\gamma_\mu C^{ab}_\mu(x) \widehat \psi^A_b(x), ~~C^{ab}_\mu=C^s_\mu
T^s_{ab}. \ee

The generating  functional can be written as

\be Z=const \int D_\mu (C) D\widehat \psi D\widehat{\bar \psi}\exp
(i \int \widehat{\bar\psi} \widehat \partial \widehat \psi d^4 x +
\int d^4 x L_{high})\label{7},\ee where $D_\mu(C)$ is the
integration over gauge field $C_\mu$ with the standard weight,
which may be also considered as the averaging over vacuum fields
$C_\mu$, denoted as  $\lan \mathcal{F} (C)\ran_C$. In this way one
can exploit the cluster expansion for $\lan \exp (\int L_{high}
d^4 x)\ran$, namely, \footnote{Note, that in general the expansion
(\ref{8}) is not gauge invariant. To make quark propagator
$S(x,y)$ and quark mass operator $M(x,y)$ gauge invariant, one
should consider quark accompanied by the parallel transporter
$\Phi(x,y)$, i.e. $tr [S(x,y) \Phi(y,x)]$. In case of confinement
this precludes definition of one-particle dynamics. Here we
consider nonconfining field $C_\mu(x)$, and then one-particle
operator $M(x,y)$ can be made gauge invariant in the local limit,
$x\to y$.}
$$ \lan \exp \int L_{high} d^4x\ran =\exp\left\{ \int J_2
(x_1,x_2) (\Psi(x_1) \Psi(x_2)) dx_1 dx_2\right.$$
$$\left.+\int J_4(x_1,.. x_4) \Psi(x_1)  ... \Psi (x_4) dx_1... dx_4
+...\right\}=$$ \be =\exp \left\{ \sum_{n=2,4,6,...} J_n (x_1,...
x_n) \Psi(x_1)... \Psi(x_n) dx_1... dx_n\right\}.\label{8}\ee Here
$\Psi(x) =\widehat{\bar \psi}(x) \gamma_\mu \widehat  \psi(x)$,
and we have suppressed spinor and group indices. Note, that only
connected correlators of field $C_\mu$ enter in $J_n$.

As a nesxt step, we do the Fierz transformation, which allows us
to form white bilinears, e.g. for two operators \be (\widehat
{\bar \psi}^A_a \gamma_\mu \psi^A_b) (\widehat{\bar \psi}_b^B
\gamma_\mu \widehat \psi^B_a)= \sum c_i (\widehat{ \bar \psi}^A_a
O_i \widehat\psi^B_a) (\widehat{ \bar \psi}^B_b O_i
\widehat\psi^A_b),\label{9}\ee where $c_i =- 1, +1,
\frac12,\frac12$ for $i=S,P,V,A,$ and  anticommutation of
operators $\widehat\psi$ is taken into account.

In a similar way one can make pairwise Fierz transformation for any $n$ in
(\ref{8}), and keeping only $S$ and $P$ terms, one arrives at the combinations
$$ \Psi (x_1) \Psi (x_2) \to (\widehat{\bar \psi} \gamma_5 \widehat \psi)  (\widehat{\bar \psi} \gamma_5 \widehat
\psi)-  (\widehat{\bar \psi} \widehat \psi) (\widehat{\bar \psi} \widehat
\psi)=$$
 \be =- \left\{  (\widehat{\bar \psi}_R \wps_L)  (\widehat{\bar
\psi}_L\wps_R) +( \widehat \psi_L\widehat \psi_R) (\widehat{\bar
\psi}_R\wps_L)\right\}\label{10}\ee with the notation $\Phi_{RL}
(x_1, x_2)\equiv\wbp_R^A(x_1) \wps^B_L(x_2)$. One can rewrite
(\ref{8}) as follows\footnote{Note, that to form white bilinears
in a connected correlator the number of fermion transmutations is
always  odd, hence, the minus sign in (\ref{11}).} $$ \lan \exp
\int L_{high} d^4x\ran_C= \exp \left\{ -\sum_{n=2,4,..} \int  J_n
(x_1,... x_n)\right.$$\be\left.\left[ X(x_1, x_2) X(x_3, x_4)...
X(x_{n-1}, x_n)\right]dx_1... dx_n\right\}, \label{11}\ee where
$X(x,y) =\Phi_{RL} (x,y) \Phi_{LR}(y,x) + \Phi_{LR} (x,y)
\Phi_{RL}(y,x).$

At this point one can do a bozonization trick, which we perform
introducing functional representation of  $\delta$- function
\cite{11*}. In short-hand notations one has\footnote{We have
suppressed the $SU(2)$ isospin subscript $i$ in $\Phi_{RLi}$ and
below in $\varphi_i, \varphi_i^+, \mu_i, \mu_i^+$; in what follows
we choose the gauge with $\varphi_1=0, \varphi_2\neq 0$.}
$$\lan \exp \int L_{high} d^4 x\ran_C= \int D\mu D\mu^+
D\varphi D\varphi^+ \exp \left\{ -i \int \mu (\varphi- \Phi_{RL})
dx- i \int \mu^+ (\varphi^+ - \Phi_{LR} ) dx \right\}\times$$
\be\times \exp \left\{-\sum_{n=1,4} \int J_n (\varphi\varphi^+ +
\varphi^+\varphi)^{n/2} dx_1 ... dx_n\right\}.\label{12}\ee Now
one can integrate over $D\wps D\wbp$ in (\ref{7}), since $\wps$
enters in (\ref{12}) only bilinearly in $\Phi_{RL}, \Phi_{LR}$. As
a result $Z$ in (\ref{7}) acquires the form

\be Z=const \int D\mu D\mu^+ D\varphi D\varphi^+
\exp(-i(\mu\varphi) - i(\mu^+ \varphi^+) + K\{\mu,
\varphi\}),\label{13}\ee where the notations are used \be S^{-1} =
i\hat
\partial + i \frac{\mu+\mu^+}{2} + i \frac{\mu-\mu^+}{2}
\gamma_5\label{14}\ee

and $$(\mu\varphi) \equiv\int \mu(x_1, x_2) \varphi(x_1, x_2) dx_1 dx_2=$$ \be
= V_4 \int \mu(x_1-x_2) \varphi (x_1-x_2) d(x_1-x_2) = V_4 \int \mu (p) \varphi
(p) \frac{d^4p}{(2\pi)^4}\label{15}\ee \be K\{ \mu, \varphi\} \equiv - \sum_n
\int J_n (\varphi\varphi^++\varphi^+\varphi)^{n/2} dx_1... dx_n+ tr ln
S^{-1}.\label{16}\ee

As a next step we find the stationary points in integration over
$D\mu D\varphi D\mu^+ D\varphi^+$

$$-i\mu (x) = \frac{-\delta K}{V_{4} \delta \varphi(x)} = \sum_n n \int
J_n \varphi^+ (x) (\varphi \varphi^+ + \varphi^+\varphi)^{n/2-1}
\delta(x_1-x_2-x) \frac{dx_1... dx_n}{V_4}$$

\be-i\varphi (x) =  tr \{ \frac{i}{2}(1+\gamma_5) S\}, ~~ -i \varphi^+ (x) = tr
\left\{ \frac{i}{2} (1-\gamma_5) S\right\}\label{17}\ee and for $\mu^+$ the
same expression, as for $\mu$, follows  with the replacement
$\varphi\leftrightarrow \varphi^+$. For the solutions of (\ref{17}) with
$\mu=\mu^+, ~~\varphi=\varphi^+$ one obtains for $\mu(p), \varphi(p)$, keeping
only terms with $n=2,4,6$,
$$ \mu(p) =\xi_2 \int \frac{d^4 q}{(2\pi)^4} J_2(q) d(p-q)
-$$
$$-\xi_4 \int \frac{d^4 q d^4q' d^4s}{(2\pi)^{12}} J_4(q,q',0) d(p-q)d(s)
d(s+q') $$ \be +\xi_6 \int \frac{d^4 qd^4q'd^4q'' d^4s d^4 s' J_6 (q, q', q'',
0,0)d(p-q) d(s) d(s+q) d(s') d(s'+q'')}{(2\pi)^{20}}. \label{18}\ee Here
$d(k)=\frac{\mu(k)}{k^2+\mu^2(k)},~~ \xi_2=\xi_4=1,~~ \xi_6=\frac34$.

Eq.(18) is the main result of this section. Solutions $\mu_i(p)$ define the
masses of different generations, $i=1,2,3,...$ and will be the subject of study
in the following sections. The composite scalar field $\varphi$ and its
nonlocal mass $\mu$ play the role of the corresponding Higgs parameters of the
standard model.

\section{Qualitative analysis of the resulting equation (\ref{18})}

At this point we specify the scales of  nonperturbative
correlators  of gauge field $C_\mu$ and denote the correlation
length of the  correlators $J_n$ as $M_n$, so that the average
value of field $\bar C_n\sim \sqrt{\lan \bar C^2_\mu\ran}\sim
\sqrt{\lan F^2_{\mu\nu}\ran M^2_n}$ ; also for simplicity we
assume that the correlation length does not depend on $n, ~M_n=M$.

Then one can introduce dimensionless quantities marked with tilde, $ \tilde \mu
(p) =\mu(p)/M,~~ \tilde p\equiv p/M, \tilde q \equiv q/M$ etc., and
dimensionless kernels $\tilde J_n$ \be J_2 =\bar C^2 M^{-4}\tilde J_2,~~ J_4=
\bar C^4M^{-12}\tilde J_4,~~ J_6 =\bar C^6 M^{-20}\tilde J_6.\label{19}\ee

As a result  Eq. (\ref{18}) keeps its form, where all quantities
are now dimensionless (with the tilde sign), and \be \xi_2 \to
\tilde \xi_2 =\xi_2 \bar C^2/M^2,~~ \xi_4 \to \tilde \xi_4 =\xi_4
\bar C^4/M^4, ~~ \tilde \xi_6 =\xi_6 \bar C^6/M^6.\label{20}\ee

Note, that the integration over momenta $\tilde p, \tilde q$ is
now over regions of the order of unity, while the mass eigenvalues
$\tilde \mu=\mu /M$ are expected to be much less than unity,
$\tilde \mu(0) \ll 1$, and $\tilde \mu(\tilde p)$ decreases with
$\tilde p$. Therefore $d(\tilde k)\cong \tilde \mu(k)/ \tilde
k^2$, and one can extract $\tilde \mu (p), \tilde \mu(q),... $
from the integrals in (\ref{18}) at some average point, $\tilde
\mu(p_*) \to \mu_*$. As a result one can approximate the integral
equation (\ref{18}) by the algebraic one: \be \mu_* =\mu_* a_2
-\mu_*^3 a_4 + \mu_*^5 a_6 - \mu_*^7 a_8 +...\label{21}\ee Here
$a_n$ are functions of $\mu_*, ~~ a_n(\mu^*) = a_n(0) + \mu^*
a_n'(0)+...$ and for $\mu_*\ll 1$ one can keep only $a_n(0)$,
which are the numbers proportional to $\left(\bar C/M\right)^{n}$.

Three solutions of (\ref{21}) are readily obtained, when $a_n=0,~~
n\geq 8$ \be \mu^2_*(1) =0, ~~\mu^2_*(2),~~ \mu^2_*(3)=\frac{ a_4
\pm \sqrt{a^2_4-4a_6(a_2-1)}}{2a_6}.\label{22}\ee To obtain $\mu_2
\ll \mu_3$ we assume that  $a^2_4 \gg 4 a_6 (a_2-1)$, obtaining in
this way \be \mu^2_* (2) \cong \frac{a_2-1}{a_4},~~ \mu^2_*(3)
\cong\frac{a_4}{a_6}.\label{23}\ee We further assume, that
$\frac{a_2-1}{a_4} =\nu^2\ll 1$ and $\frac{a_4}{a_6} \cong\nu$.
Then masses of second and third generations are \be \mu(2) = \mu_*
(2) M \cong \nu M,~~ \mu(3) \cong \sqrt{\nu} M,~~ \nu\approx
\left(\frac{\mu(2)}{\mu(3)}\right)^2.\label{24}\ee From
experimental quark masses one has that $\nu\approx 10^{-4}$ for
$(u,c,t)$ and $\nu\approx 10^{-3}$ for ($d,s,b)$; parametrically
$\nu\sim \left( \frac{M}{C}\right)^2,$ and $M$ from (\ref{24})
turns out to be $M\sim \frac{m^2_3}{m_2}\sim 20 $ TeV from the
$(u,c,t)$ sector.

To make connection with previous results in topcondensate-type
models [3,4,5,6], one should neglect all $a_n$ except
$a_2(\mu^*)$, and returning to unscaled variables and identifying
flavor-depending mass $M^{AB}=\mu^{AB}$, writes

\be \mu^{AB}=\frac{\bar g^2}{M^2}\int \frac{d^4p}{(2\pi)^4}
\mu^{AC} (\frac{1}{p^2+\mu^2})_{CB}\label{25} \ee

Here $\bar g^2 \equiv \xi\left( \frac{\bar C}{M}\right)^2$ estimates the kernel
$J_2$, and   cut-off at $p\sim M$  is assumed.

One can see, that Eq.(\ref{25}) is easily diagonalized in the
flavor-democratic manner  and one is facing the familiar
fine-tuning problem \cite{4,5}, where $\bar g^2 = \bar g^2_{cr}
+O\left(m^2_0/M^2\right)$. The ways to circumvent this problem are
suggested in TC, ETC and TC2 models (see \cite{6,12} for a
comprehensive review). However, in this Gaussian approximation
(when only $a_2$ is kept nonzero) it is not clear how to get three
generations with distinct masses, different from flavor democratic
scenario \cite{8,9,10}.

As was shown above this can be done keeping nonzero three
coefficients: $a_2, a_4, a_6$ and this allows to obtain three
generations with masses, which can be made much different by an
appropriate choice of the coefficients $a_2, a_4, a_6$. However,
here the first generation acquires zero masses and to obtain
realistic values of the masses new mechanism will be introduced in
next Sections. A  negative feature of the result (\ref{23}) for
$\mu(3)$  is that it depends only on the field correlators via
$a_4, a_6$ etc. and at this point two questions arise: why $a_4,
a_6 $ etc. should be so much different, since the expansion
parameter $\left(\frac{\bar C}{M}\right)^2$ cannot be too large
for realistic field configurations; also it is not clear why
$\mu(3) $ is so much different for $b$ and $t$ quarks. In Section
5 we shall introduce a  mechanism, which can in principle explain
this high sensitivity  of the quark masses and appearence of the
realistic hierarchy of masses even if the coefficients $a_n$ are
of the same order of magnitude.

\section{Topological charges in the  EW vacuum}

In this  section we demonstrate, that an admixture of topcharges in the EW
vacuum can drastically change the masses of the lowest generation.

At this point we would like to see the  effects of topological charges
(topcharges). We do not specify here the character of topcharges and its group
assignment, assuming that in nonperturbative vacuum of field $C_\mu$ there are
group-topological conditions for existence of  corresponding solutions, similar
to $SU(2)$ instanton solutions. Therefore the fields of topcharges $A_\mu^{(i)}
(x-R_i)$ in singular gauge, located at points $R_i$, have to be added to the
fields \be C_\mu(x) \to C_\mu(x) + \sum^N_{i=1}
A^{(i)}_\mu(x-R_i).\label{26}\ee In this case the generating functional
includes averaging over topcharge positions, sizes and orientations, denoted as
$D\Omega$, \be Z=const \int D\kappa (C) D\Omega D\wps D\wbp\exp\{
\mathcal{L}\}\label{27}\ee where $\mathcal{L}= \mathcal{L}_1 + \mathcal{L}_2+
\mathcal{L}_3$, and

\be\left.\begin{array}{l} \mathcal{L}_1 =i\int \wbp_f \hat \partial \hat \psi_f
d^4
x ; ~~\mathcal{L}_2 = \int \wbp_f C_\mu \gamma_\mu \wps_fd^4x,\\
\mathcal{L}_3 = \sum^N_{i=1}\int \wbp_f \hat A^{(i)} (x-R_i) \wps_f
d^4x\end{array}\right\}\label{28}\ee Now we average over the fields  $C_\mu$ as
before, keeping topcharges intact, which brings in  a new term, estimated at
the stationary points,  taking $\mu=\mu^+$, one has   \be \mathcal{L}_2 \to
\int \wbp_f (x) i \mu(x,y) \psi_f (y) d^4x d^4y.\label{29}\ee As a next step we
average over topcharges in the same way, as it was done in case of instantons
in \cite{13,14,15}, see also Appendix 1. This yields the following quark
Green's function (see Appendix 1 for derivation). \be S(x,y) = S_0 (x,y) +
\sum^\infty_{s=1} u_s (x) \frac{1}{\Lambda_s-i\mu} u^+_s (y),\label{30}\ee
where $u_s, \Lambda_s$ are the eigenfunction and the eigenvalue of the ensemble
of topcharges, and $S_0 = (-i\hat
\partial -i\mu)^{-1}$ and $\mathcal{L}= -\int \wbp S^{-1} \wps d^4 x.$

Note, that in general the sum in (\ref{30}) contains $N_0$ zero modes,
corresponding to the net topcharge density $\frac{N_0}{V_4}$, and a region of
quasizero modes. Neglecting correlations between topcharges, the eigenvalues
follow the Wigner semicircle law in case of instantons \cite{14,15}. Going to
the momentum space and averaging over topcharge positions, one has \be S(p)
=\frac{1}{\hat p-i\mu} + \frac{c_0(p)}{-i\mu}+ \frac{c_1(p)}{\bar
\mu-i\mu},\label{31}\ee where we have separated contributions of zero and
quasizero modes and defined \be c_0(p) =
\frac{N_0|\overline{u_0(p)}|^2}{V_4},~~c_1(p) =
\frac{N|\overline{u_1(p)}|^2}{V_4}, ~~\bar
\mu=O\left(\left(\frac{\rho}{R}\right)^4\right).\label{32}\ee Here $u_0(p),
u_1(p)$ are zero and quasizero  eigenfunctions; $\rho$ is the average topcharge
size, while $R=\left(\frac{V_4}{N}\right)^{1/4}$ is the mean distance between
topcharges.

To understand the change in the basic Eq.(\ref{18}) due to
topcharges and introducing $d(p)\equiv \frac{\mu(p)}{D(p)}=
\frac{\mu(p)}{p^2+\mu^2(p)}$, one should compute  $$
\frac{\partial tr}{\partial \mu} ln S^{-1} = - tr \left(
\frac{1}{S} \frac{\partial S}{\partial\mu}\right).$$ Firstly, we
simplify (\ref{31}) and take  $c_1(p)\equiv 0$. In this case one
has \be d_{top} (p) =\frac{\partial}{\partial\mu} tr ln S^{-1} =
\frac{\mu}{p^2+\mu^2}+ \frac{(c_0p)^2}{(c_0 p)^2 +
\mu^2}.\label{33}\ee

Since in (\ref{33}) $\lan p\ran\sim M$, and $\mu^*=\mu/M$,  for
the $d$-factor in (\ref{18}) in case of no topological charges $
one has d_*= dM\approx \mu^*$, which produces  Eq.(\ref{21}). In
case, when topological charges are present, i.e. $(c_0\neq 0)$,
one has \be d_* (topcharge) \equiv \bar d\cong
\frac{\mu_*}{1+\mu_*^2}+ \frac{c^2_0}{(c_0^2+\mu^2_*)
\mu_*}\approx \mu_* +
\frac{c_0^2}{(c_0^2+\mu^2_*)\mu_*}\label{34}\ee and Eq.(\ref{21})
can be rewritten as \be \mu_* = a_2 \bar d - a_4 \bar d^3 + a_6
\bar d^5-...\label{35}\ee

It is easy to see,  that for $\mu_* \gg \sqrt{c_0}$  one recovers
old result: $\bar d\to  d_* \approx \mu_*$, with two roots given
in (\ref{23}). However, the zero solution for $\mu_*$, valid for
Eq. (\ref{21}), is not possible in Eq. (\ref{35}). Instead, for
$\mu_*\la \sqrt{c_0},$ neglecting the terms $a_n, n \geq 4$ in
(\ref{35}), one obtains one root, \be \mu^2_*(1) \approx c_0
\left(\frac{ a_2}{1-a_2}\right)^{1/2},~~ m_1\approx \mu_* M\approx
\sqrt{c_0}M\left(\frac{ a_2}{1-a_2}\right)^{1/4}.\label{36}\ee

Hence, the mass of the lowest generation is defined by the vacuum admixture of
unbalanced zero models, i.e. by the topological charge of the vacuum.  It is
clear that higher order terms with $a_4, a_6$,... contribute corrections to
$\mu_*(1)$, which are of the order of $c_0^{3/2}, c_0^{5/2}$ etc. and can be
neglected. Thus in case of topcharges one obtains a new (and nonzero) solution
$\mu_*(1)$, while the masses of higher generations are kept intact.

In the system (\ref{35}) there might be other solutions, in addition to three
solutions (\ref{36}) and (\ref{23}), which are of the order of
$M\sqrt{\frac{a_4}{{a_2}}}, M\sqrt{\frac{a_6}{{a_4}}}$ etc. However, with
discussed above assumptions, $\frac{a_2-1}{a_4}\ll 1, ~\frac{a_4}{a_6} \ll 1$,
these roots appear to be higher than the scale $M$ and therefore unphysical.
Other and smaller roots could appear, if $\frac{a_4}{a_2} \ll c_0,
~~\frac{a_6}{a_4} \ll c_0$ etc., which do not seem reasonable for realistic
$c_0 \la 10^{-8}$.

One more general remark is in order. As shown in Appendix 1, the approximation
used in (\ref{30}), when $\mu$ enters in the denominator of second term, is
valid when the correlation length of the vacuum $M^{-1}$ is much smaller than
the size of topcharge $\rho$, $ M^{-1}\ll \rho$. Then one can consider
(nonlocal) effective mass $\mu$ as a constant inside the topcharge field. It is
clear, that in general the high-n correlator term $J_n(x_1,... x_n)$ will be
outside of topcharge size $\rho$ for high enough $n$, hence, zero mode
contribution (second term in (\ref{34})) will be effectively damped in high-n
terms $a_n \bar d^n\to a_n \mu^n_*$. Therefore one cannot expect additional
roots from higher $a_n\bar d^n$ terms, which are not present in the
$a_n\mu_*^n$ series  (\ref{21}). Hence we except, that topcharges can create
small masses of the lowest generation, with the scale proportional to the
concentration of topcharges.

\section{Mixing due to fermion bilinear condensation. The mechanism of coherent mixing}

Till now we have disregarded the matrix nature of  $\mu_{ik},
\varphi_{ik}$. As it is clear from  (\ref{9})-(\ref{12}),
$\Phi_{RL}(x_1, x_2) \equiv \wbp^A_R (x_1) \wps^B_L(x_2)$ is a
matrix in the indices $A=(n,\alpha), ~~ B= (n',\beta)$, where
$n,n'$ refer to families and $\alpha,\beta$ to fermion sectors. As
will be shown below,the sector mixing in $\alpha,\beta$ does not
occur from the effective Lagrangian $K\{ \mu, \varphi\},$ Eq.
(\ref{16}), but the family mixing does occur. For notational
convenience in the matrices $\hat \mu_{nn'}, \varphi_{nn'}$,
instead of $n,n'$ we shall denote the family numbers $i,k$ as
$\mu_{ik},~~ i\hat\varphi_{ik},~~ i,k=1,2,3,...,$ and  use the
relation\be \mu_{ik} (p) = (p^2+ \hat \mu^2) _{il} \varphi_{lk}
(p),~~ d_{ik} (p) =\hat\varphi_{ik} (p).\label{37}\ee Then  the
basic equation (\ref{18}) can be rewritten in the $x$-space as
(for $|\hat \mu^2(p)|\ll p^2),$
\be-^{\framebox{}}~\widehat{\varphi}(x) =- \frac{\delta
U\{\widehat{\varphi}\}}{\delta \widehat{\varphi} (x)} = \tilde J_2
\widehat{\varphi} -\tilde J_4\widehat{\varphi}^3 +\tilde
J_6\widehat{\varphi}^5-...,\label{38}\ee where $\tilde J_n$ are
actually nonlocal kernels  in $x$-space corresponding to the
$p$-space kernels in ({18}). Also the functional
$U\{\widehat{\varphi}\}$ can be written as \be
U\{\widehat{\varphi}\}= - \frac12 \tilde J_2 \widehat{\varphi}^2+
\frac14 \tilde J_4 \widehat{\varphi}^4-\frac16  \tilde J_6
\widehat{\varphi}^6+...\label{39}\ee
Our functional $U\{\hat\varphi\}$, Eq. (\ref{39}) has the standard
form which was already investigated in the local limit in search
for solitonic solutions \cite{16,17}. As was discussed above,
$U\{\widehat{\varphi}\}$ has several stationary points, which we
associate with the $p$-space solutions of Eq. (\ref{18}) or the
$x$-space solutions,  of Eq. (\ref{38}). These  solutions are not
like solitons, but rather solutions of nonlinear nonlocal integral
(or integro-differential) equation, where the values of
$\varphi_{ik}$ are varying in the region around the stationary
point $\hat\varphi_n (x)$ of $U\{\hat\varphi\}$. Therefore one can
identify the diagonal elements of the matrix function
$\widehat{\varphi}(x)$ with $\varphi_n (x)$ as follows
\be\hat\varphi_{ik} (x) = \delta_{in} \delta_{kn} \varphi_n(x), ~~
n=1,2,3,...\label{40}\ee

Let us now turn to nondiagonal elements of $\hat\varphi_{ik}(x)$.
>From physical point of view, since the diagonal elements are
associated with stationary points of $U\{\widehat{\varphi}\},$
nondiagonal elements $\hat\varphi_{ij}$ should be solutions
connecting two stationary points $i,j$, i.e. solutions of kink
type. A well-known example of  the kink for the functional
$U(\hat\varphi) = \frac{\lambda}{4} (\hat\varphi^2-m^2/\lambda)^2
$ is given by the solution $\hat\varphi_{kink} (x) =
\frac{m}{\sqrt{\lambda}} th \left( \frac{m}{\sqrt{2}}
(x-x_0)\right)$, which connects stationary points $\hat\varphi_1 =
\frac{m}{\sqrt{\lambda}}$ and $\hat\varphi_2=
-\frac{m}{\sqrt{\lambda}}$. Therefore we shall assume here, that
similar solutions exist in  our case for Eq. (\ref{38}) with the
value of $\hat\varphi_{ik}(x)(i\neq k)$ varying in the region
between stationary "points" $\varphi_n (x)$ with $n=i$ and $n=k$.

Now coming back to qualitative discussion (in Section 3) of
possible solutions $\mu(p)$ of Eq. (\ref{18}) or
$\hat\varphi_{ik}(x)$ of Eq.(\ref{38}), one expects the average
fermion masses $\mu_{ik} (p) \to \mu_{ik}$ to be equal to $\mu(n)$
for $i=k=n$, and  some average of $\mu(i)$ and $\mu(k)$ for $i\neq
k$. In what follows we take $\mu_{ik}$ for the kink solution as
the "geometrical average" \be \mu_{ik}=\sqrt{\mu(i) \mu(k)}
\label{41}\ee and in this case the spectrum with all eigenvalues,
with an exception of the largest one, appear to be arbitrarily
small. We call  this phenomenon  the Coherent Mixing Mechanism
(CMM).

To study qualitatively the CMM in more simple case of two
families, we start with the mass matrix $\mu_{ik} $ both in up and
down sectors: \be \mu_{ik}= \left(
\begin{array}{ll} \mu_1&
\mu_{12}\\\mu_{12}&\mu_2\end{array}\right).\label{42}\ee The
eigenvalues of $\mu_{ik} $  in the case with $\mu_2\gg \mu_1$ are
\be m_+ = \mu_2 +\frac{\mu^2_{12}}{\mu_2}, ~~ m_- = \mu_1
-\frac{\mu^2_{12}}{\mu_2}.\label{43}\ee For the choice (\ref{41})
and  $\mu_{12}= \sqrt{\mu_1\mu_2}$, one has $m_-=0,~~
m_+=\mu_1+\mu_2$ and the CKM matrix $V_{ik}$ has the form \be \hat
V= W_u  W_d^+= \left(\begin{array}{ll}
1-\frac{\bar\eta^2}{2}&\bar\eta\\
-\bar\eta& 1-\frac{\bar\eta^2}{2}\end{array}\right)\label{44}\ee
with $\bar \eta\cong\sqrt{\frac{\mu_1}{\mu_1+\mu_2}}$, which refer
to the sector $(d,s,b)$, since $(u,c,t)$ sector yields much
smaller $\eta$). It is clear, that varying $\mu_{12}$ around the
value $\sqrt{\mu_1\mu_2}$, one obtains physical values of $m_1$ in
the region from $\mu_1$ to zero, and hence $\mu_1$ can be $\mu_1
\ga m_d$, while $ \mu_2 \approx m_+ \sim m_s$, and $\eta$ has a
reasonable value , $\bar\eta\approx 0.2$, in agreement with
experimental data. Note, that in CMM the choice (\ref{41}) gives
the maximal value of mixing for fixed $\mu_1, \mu_2$ and minimal
value of $m_-=0$.

For three or more generations one can write the general CMM matrix
$\hat\mu $ as follows\footnote{Our input mass matrix contains
$\mu_i,\mu_{ik}$, and the final diagonalized mass matrix contains
physical masses $m_i, i=1,2,3~~(m_u, m_c, m_t$ and $m_d, m_s,
m_b$).}
\be \hat\mu =\left(\begin{array}{cccc} \mu_1& \mu_{12}&\mu_{13}&...\\
\mu_{12}& \mu_2& \mu_{23}&...\\
\mu_{13}&\mu_{23}& \mu_3&...\\
...&...&...&...\end{array}\right),~~ \mu^2_{ik}= \mu_i\mu_k
(1+\eta_{ik})\label{45}.\ee

Here  we denote a diagonal effective mass $\mu_{ii}$ as $\mu_i$;
it corresponds to the i-th minimum of the functional
$U(\widehat\varphi),$ and then consider $\eta_{ik}$ small,
$|\eta_{ik}|\ll 1$\footnote{ Note, that for $\mu_i =\mu_0,~~
(i=1,2,3)$ and  $\eta_{ik}=0$ one recovers the "ultrademocratic"
mass matrix \cite{7}  -\cite{9}.}.

We assume that $\mu_1< \mu_2<\mu_3<... $ and $\mu_i$ not
necessarily much smaller than $\mu_{i+1}$.

The eigenvalue equation in case of tree generations, $\det (\hat
\mu-\hat m I)=0$, given in  Appendix 2, Eq.(\ref{A2.3}), can be
written as \be m^3 - m^2 \sigma + m\xi -\zeta =0\label{46}\ee
where coefficients can be expanded in powers of $\eta_{ik} $,
yielding  $\sigma=\sum_i\mu_i,$ $ \xi = - \sum_{i\neq j} \mu_i
\mu_j \eta_{ij}, $ $  \zeta =\mu_1\mu_2\mu_3\left(-\frac14
\sum_{i\neq j} \eta^2_{ij}+  \frac12 \sum_{i\neq j~ l\neq k}
\eta_{ij} \eta_{lk}\right)$.  Since  $m_1m_2 m_3 =\zeta$, for very
small $\zeta$ the lowest eigenvalue $m_1$ tends to zero, and one
obtains the following hierarchy of eigenvalues \be m_1 \cong
\frac{\zeta}{\xi},~~ m_2 \approx \frac{\xi}{\sigma}, ~~ m_3
\approx \sigma.\label{47}\ee Note, that one can easily adjust the
strong hierarchy, namely,  $m_1 \ll m_2 \ll m_3$ by a simple
variation  of the parameters $\eta_{ij} $. For example, as shown
in Appendix 2  for the  choice $\eta_{23} = \eta_{13}
\equiv-\eta,~~ \eta_{12} =-\delta, ~~ 0<\delta\ll \eta$, one has
\be m_1 \approx \mu_1 \delta, ~~m_2\approx \frac{\mu_3}{m_3}
(\mu_1+\mu_2)   \eta, ~~ m_3 \approx
\mu_1+\mu_2+\mu_3.\label{43}\ee It is interesting, that for
$\delta \ll \eta \ll 1 $ the CMM has made the hierarchy much more
pronounced, than original situation with $\mu_1< \mu_2< \mu_3$,
and $m_1, m_2$ can be made very close to zero,  while $m_3$ is not
far from $\mu_3$.

The mass matrix (\ref{45}) is diagonalized, as shown in Appendix 2, with the
help of the unitary matrix $W$, Eq.(\ref{A2.13a}), where we also introduced
imaginary parts in $\mu_{ik}$ as $\mu_{ik} = |\mu_{ik}| e^{i\delta_{ik}}$ to
account for a possible CP violation with the condition
$\delta_{12}-\delta_{13}+\delta_{23}=0$.  A very convenient way of constructing
unitary matrices $W$ in terms of $\mu_{ik}$ and $m_i$ was given in \cite{17}.
There a simplified form was obtained in case $W_{13} \to 0$. In our specific
case of CMM matrix (\ref{45}), the element $W_{31}$ tends to zero when
$\delta\to 0$, and another  simple form for $W$, shown in (\ref{A2.13}),
occurs. The resulting CKM matrix $\hat V$ is readily computed from $W_u$ and
$W_d$ Eq.(\ref{A2.13}) and it is given in the Appendix 2, Eq.(\ref{A2.17}). It
is expressed  in terms of the phases $\delta_{ik}$ and the values $m_i, \mu_i$
only. The latter enter via sines and cosines  defined in (\ref{A2.16a}).

Using (A2.17) we can write a simplified version of CKM matrix realizing, that
all cosine factors are equal  to unity within $(1\div 2)\%$, while the sine
factors $\sim 0.1$, and one has the following estimates from PDG \cite{18}

\be |V_{ud}| = |c^u_\alpha c^d_\alpha + s^u_\alpha s^d_\alpha
e^{i\Delta_{12}}| = 0.97418\pm 0.00027\label{51}\ee

\be |V_{us}|\cong |s^u_\alpha  e^{i\delta^u_{12}}-s^d_\alpha
 e^{i\delta_{12}^d}| = 0.2255\pm
0.0019\label{52}\ee

\be |V_{cd}|\cong |s^d_\alpha  e^{-i\delta^d_{12}}-s^u_\alpha
 e^{-i\delta_{12}^u}| = 0.230\pm
0.011\label{53}\ee

 \be |V_{cs}|=|c^u_\alpha c^u_\beta c^d_\alpha c^d_\beta +O(ss)| = 1.04\pm
0.06\label{54}\ee

\be |V_{tb}|= |c^u_\beta c^d_\beta +O(ss)| >0.74 \label{55}\ee

\be |V_{ub}|= |s^u_\alpha (s^u_\beta  e^{i\delta^u_{13}}- s^d_\beta
e^{i(\delta^u_{12}+ \delta^d_{23})})|= (3.93\pm 0.36) {10^{-3}}\label{56}\ee

\be |V_{cb}|= |s^u_\beta e^{i\delta^u_{23}}- s^d_\beta
e^{i\delta^d_{23}}| =(4.12\pm 0.11) {10^{-2}}\label{57}\ee

\be |V_{ts}|= |s^d_\beta e^{-i\delta^d_{23}}- s^u_\beta
e^{-i\delta^u_{23}}| =(3.87\pm 0.23)  {10^{-2}}\label{58}\ee

\be |V_{td}|= |s^d_\alpha (s^d_\beta e^{-i\delta^d_{13}}- s^u_\beta
e^{-i(\delta^u_{23}+\delta^d_{12})})|= (8.1\pm 0.6) {10^{-3}}.\label{59}\ee

One can see, that two equalities arise from the above expressions
\be V_{us} = - V^*_{cd}, ~~ V_{ts}=- V_{cb}^*\label{60}\ee

The experimental values for moduli of these expressions are equal
within the errors.

The angles $\alpha, \beta,\gamma$ are easily computed from the entries of
(\ref{51})-(\ref{59}), namely: \be \alpha=\Delta_{12}, ~~ \beta= \arg
(s^d_\alpha-s^u_\alpha e^{-i\Delta_{12}}),~~\gamma=\pi-\Delta_{12}
-\beta.\label{59a}\ee Assuming $\alpha\approx \frac{\pi}{2}$ \cite{18}, our
prediction is $\beta \cong 24^\circ$, yielding $\sin 2 \beta \cong 0.73$ not
far from experiment.

Moreover, dividing (\ref{56}) by (\ref{58}) and using the
condition $\delta_{13}-\delta_{12}=\delta_{23}$, one obtains
$s^u_\alpha \cong 0.1$, which yields $\mu_u \approx 20$ MeV,
whereas dividing $V_{td} $ (\ref{59}) by (\ref{58}), one obtains
$s_\alpha^d\approx 0.21$, which defines $\mu_d\approx 10$ MeV
(both at the scale of 2 GeV). An important check is the
reparametrization -invariant quantity $ \bar \rho+ i\bar \eta=
-\frac{V_{ud}V^*_{ub}}{V_{cd} V^*_{cb}}$, which in our simplified
ansatz (\ref{A2.13}) is

$$ \bar \rho+ i\bar \eta= \frac{s^u_\alpha c^d_\alpha (s^u_\alpha - s^d_\alpha
e^{-i\Delta_{12}})}{(s^u_\alpha)^2+ (s^d_\alpha)^2}\approx 0.18+ i 0.38.$$

The assumption $\Delta_{23}\equiv 0$ makes possible a strong
cancellation in $(s^d_\beta -s^u_\beta e^{-\Delta_{23}})\approx
(s^d_\beta -s^u_\beta)$ which should be as small as 0.04, yielding
$\mu_s \approx 0.35 $ GeV, $\mu_c \approx 14$ GeV. Using
(\ref{51})-(\ref{59}) one has a reasonable estimate for $J$, Eqs.
(\ref{A2.18}), (\ref{A2.19}), if $s^d_\beta$ and  $s^u_\beta$
satisfy (\ref{58}),
$$J=s^u_\alpha s^d_\alpha |s^d_\beta -s^u_\beta e^{i\Delta_{23}}|^2\sin \Delta_{12}\approx 3.2\cdot 10^{-5}\sin
\Delta_{12},$$ which agrees well with the PDG value \cite{18}.
 Thus all phenomenological tests are passed by our
representation.

 At this point one could estimate more systematically  the input values
of $\mu_{ik}$ necessary to satisfy experimental data for $V_{ik}$.
We shall not do it here and in the next Section we present the
arguments, that this procedure is better to be used in case of
four generations.

In Appendix 3 we explain that in CMM the eigenvalues of four
generations obey the same pattern as in the case of three
generations - with a steady highest mass and  volatile lower
masses.

\section{Spontaneous symmetry breaking, the Higgs condensate, and the  fourth
generation}

The fermion bilinear condensation discussed above plays the same
role in the EW spontaneous symmetry breaking, as the standard
Higgs mechanism, and in this sense is an extension of the
topcondensate mechanism to the case of many generations. Since
there are many condensates , $ \lan \hat\varphi_{ij}\ran$ and
$\lan\hat \varphi^+_{ij}\ran$, one might worry about multiple
composite Higgs bosons and in this Section we shall study the
situation with the Higgs condensate (and hence with the $W$ and
$Z$ masses) and the scalar excitation of the condensate ("the mass
of Higgs boson").

To obtain the Higgs condensate, one can use the Pagels-Stokar
relation \cite{20}, where the whole spectrum of quarks in the
quark loop diagrams for the scalar current is introduced. In the
leading order it gives \be v^2 = \frac{N_c}{4\pi^2} \sum_i
\mu^2(i) \ln \frac{M}{\mu(i)},\label{49}\ee where $v=246$ GeV is
the standard value of the Higgs condensate, and the sum is over
all quark masses $\mu(i)$ in $n$ generations. With the hierarchy
already known for three generations and assumed for four
generations, the dominant contribution comes from the highest mass
$\max \mu(i)=\mu_{\max}$, and from (\ref{49}) $\mu_{\max}$ for a
given $M$ can be estimated. The results are given in the Table.
\\

For $M_W$ and $M_Z$ one has the standard relations \be M^2_W \cong
\frac{g^2_2}{4} v^2,~~ M^2_Z\cos^2 \Theta_W \cong \frac{g^2_2}{4}
v^2.\label{50}\ee

Table. The values of highest quark masses for different  mass
scales $M$ from Eq.(\ref{49})- in  second row, and from
\cite{20*}--in  bottom row.\\

\begin{tabular}{|l|l|l|l|l|l|l|l|}
\hline M&$10^{19}$&$10^{17}$&$10^{13}$&$10^{9}$&$10^{5}$&$10^{4}$&$5\cdot
10^{3}$\\
GeV&&&&&&&\\ \hline $\mu_{\max}$& 143&153&180&228&377&518&616\\
GeV&&&&&&&\\ \hline$\mu_{\max}$& 253&259&279&316&446&591&-\\
GeV&&&&&&&\\ \hline
\end{tabular}

\vspace{1cm}

Note, that (\ref{49}) is a naive approximation with  $\mu(i)$ as a
constant, which does not depend on momenta of loop integration.
Taking that into account one obtains much higher values of
$\mu_{\max}$ for a given $M$ \cite{20*} ( see  Table), hence,
$\mu_{\max}$ cannot be associated with the top quark mass (at
least in this approximation). For more discussion see the review
by Cveti\v{c} \cite{5} and Refs. therein.

For $M$ smaller than $5\cdot 10^3$ GeV the Eq.(\ref{49}) is not a
good approximation and one should use a function $\mu_i(p)$ as in
Eq. (\ref{18}), however, the consistency of the whole approach is
questionable for $\mu_{\max}\sim M$. Therefore  we onsider
$\mu_{\max}\sim 0.5 \div 0.6$ TeV as the maximal value of the $t'$
mass, and it is clear that $t'$ mass should belong to the 4th
generation of quarks.

As we noted in Section 5, the CMM ensures  volatility of the
masses of all lower generations, while for the highest generation
the masses are rather stable. This picture is consistent with what
should happen to the masses of the 3d generation, when there
exists the fourth generation. Indeed, $m_b \approx$ $ O(4$ GeV)
and $m_t\approx O(180$ GeV) are very different, and might be seen
as a subject to large changes, when small changes in the mixing
coefficients $\eta_{ik}$ in Eq.(\ref{45}) are made. In this way
assuming existence of four generations one might resolve the old
problem in the top condensate mechanism.

It is remarkable, that the analysis of precision data in \cite{21} for $m_{b'}
=m_{t'}=300$ GeV and the mass of heavy charged lepton $m_E=200$ GeV shows the
same $\chi^2$ minimum for four generations as for three, supporting in this way
the idea of the fourth generation. Note also, that for degenerate $b'$ and $t'$
quarks the mass of each of them is roughly by $(35\div 40\%)$ less than shown
in the Table. For discussion of the possible parameter space of the fourth
generation see \cite{23}.

Let us now discuss the topic of the possible (composite) Higgs
boson mass. It is clear, that the induced Higgs Lagrangian should
have the same form as in the topcondensate case,  where the Higgs
field $h$ is the deflection from the stationary point $i=j$ of the
effective potential $h=H-v\equiv \mu_{ii} -v$ \be V(H) =- m^2_H
H^+H + \frac{\gamma_4}{2} (H^+H)^2+ \frac{\gamma_6}{3}
(H^+H)^3+...-(\bar \psi_{Li} \psi_{Ri} H+ h.c).\label{61}\ee

The  difference from the standard topcondensate case is that i)
there  only one minimum of $V(H)$ exists at $H=v$, while in our
case at least three minima should be present; ii) higher order
terms in $h$ are present already at the tree level, e.g. $\gamma_4
(tree) \neq 0$ due to higher correlator terms $J_n, n\geq4$ ,
while for the topcondensate case (only $J_2 $ present) these terms
are induced by fermion loop diagrams. Moreover the higher
correlator terms contribute dominantly to the coefficient of
$h^2/2$ (the Higgs mass) in the situation discussed in Section 3
(e.g. $a_6\gg a_4>a_2$ etc.). Hence, we can conclude  that in this
case the   lowest  Higgs mass appears for the largest minimum,
i.e. near $\mu_{ii} =\mu_{\max}$. Then (also neglecting admixture
of lower minima) all coefficients are mostly quark loop-induced
and hence, $m_{Higgs} \cong 2 \mu_{\max}$, as well known
\cite{20}, \cite{4}. In our favored case of four generatons, it
means that the Higgs mass is around 1 TeV (for $M\sim 10^4 -10^5$
GeV). As noted in \cite{21}, this situation of high Higgs mass and
one extra generation does not contradict precision data.

One should also have in mind, that this Higgs boson is not
elementary and, if it exists at all, can be associated with an
excited unstable $q\bar q$ state. For more discussion of the new
physics with Higgs and  the fourth generation see \cite{22}.

\section{Conclusions and outlook}

Results of the  paper are threefold. First of all, we present a
possible dynamical scheme of $q\bar q$ pair condensation which
might explain the generation structure of the fermion hierarchy.

Secondly, we have found in the same scheme a possible source of fermion mixing,
identifying it with the kink-type (soliton-type) solutions of the same
effective potential.

Finally, we have suggested a new type of mixing pattern, called
the coherent mixing mechanism (CMM), which follows from two first
solutions. In this way one obtains a simple parametrization of the
CKM matrix $\hat V$: in terms of two phases and the input masses
$\mu_i$ (corresponding to minima of effective potential), and
resulting physical masses  $m_i$.

In CMM strong mass shifts $(\mu_i-m_i)$ of all lower generations
are caused by tiny changes in mixing parameters, which may explain
large mass differences in $u$ and $d$ sectors. The same volatility
in the masses for the third generation  and experimental values of
$V_{ik}$ strongly prefer the scheme of four generations with
$m_{t'}\sim m_{b'}\sim O(300\div 500$ GeV)\footnote{Four
generations were found favorable also in the context of flavor
democracy scenario \cite{24}.}. In this case also the Higgs
condensate and $W, Z$ masses are correctly reproduced. The method
of the paper allows to predict more explicitly the resulting
masses $m_{t;}, m_{b'}$ and some mixing coefficients, which is
planned for future publications.

As an additional possibility the role of topcharges in the EW
vacuum is studied and shown to be effective in producing very
small masses of the first generation; also the origin of the
CP-violating phases can be directly connected to the topcharge
contents of the EW and GUT vacuum, as was suggested before in
\cite{25} in another framework.

\section{Acknowledgements}

The author is grateful to A.M.Badalian for careful reading of the
manuscript and many useful suggestions, V.A.Novikov and
V.I.Shevchenko for discussions; the financial help of the RFBR
grant 09-02-00629a, and NSh-4961, 2008.2 is gratefully
acknowledged.

\vspace{2cm}
 \setcounter{equation}{0}
\renewcommand{\theequation}{A1.\arabic{equation}}

{\bf \large
\noindent Appendix 1}\\

\noindent{\bf \large The quark Greens's function in presence of
topological charges }\\

One starts with the gauge field of the form\be
 C_\nu(x) = B_\nu (x) + \sum^N_{i=1} A_\nu^{(i)}(x)\label{A.1}\ee
 where $B_\nu$ are nontopological fields, while $A_\nu^{(i)} (x)$
 are fields of topcharges, exact form of those is not important
 for us, but we shall assume that topcharges form a dilute gas
 with no correlations. Our goal is the calculation of the full
 quark Green's function $S$ in the field $C_\nu(x)$,
 \be S=\left( - i\hat \partial - g \hat B - g \sum^N_{i=1} \hat
 A^{(i)}\right)^{-1}\label{A.2}\ee
 in terms of individual topcharge Green's functions $ S_i =(-i\hat
 \partial-gB-g\hat A^{(i)})^{-1}$ and $S_0 = (-i\hat
 \partial-g\hat B) ^{-1}$.

 One can use the same technic as exploited in the Faddeev-type
 decomposition of the Green's function for particle scattering on
 many centers \cite{13}. Introducing $t$-matrices, $t_i\equiv S_0-S_i$
 and amplitudes for scattering $Q_{ik}$, where $i$ refers to  the
 first and $k$ -- to the last scattering center, one has
 equations for $Q_{ik}$
 \be Q_{ik}= t_i \delta_{ik} - t_i S^{-1}_0\sum_{j\neq
 i}Q_{jk}.\label{A.3}\ee

For $S_i$ one can use the spectral representation \be S_i (x,y)= \sum_n
\frac{u_n^{(i)}(x) u_n^{(i)+}(y)}{\mu_n^{(i)} -im},\label{A.4}\ee where we have
introduced the quark mass $m$ for future convenience, and $u_n^{(i)} (x)$
satisfies equation \be -\gamma_\mu (\partial_\mu-ig B_\mu-ig A_\mu^{(i)}(x))
u^{(i)}_n (x) = \mu^{(i)}_n u_n^{(i)}(x).\label{A.5}\ee Using (\ref{A.4}) one
can rewrite (\ref{A.3}) as follows
$$ Q_{ik} (x,y) =\delta_{ik} \left[ S_0 (x,y) - \sum_n\frac{u_n^{(i)}(x) u_n^{(i)+}(y)}{\mu_n^{(i)}
-im}\right]+$$ \be +\int d^4 z\sum_n \frac{u_n^{(i)}(x)
u_n^{(i)+}(z)}{\mu_n^{(i)} -im} (-i \hat \partial - g\hat B - \mu_n^{(i)})
\sum_{j\neq i }Q_{jk}(z,y).\label{A.6}\ee Solving (\ref{A.6}) for $Q_{ik}$, one
immediately finds $S$, \be S=S_0 - \sum^{N}_{i,k=1} Q_{ik}.\label{A.7}\ee

One can represent $Q_{ik}$ as follows \be Q_{ik} (x,y) = \sum_{n,n'}
\frac{u_n^{(i)}(x) R_{nn'}^{ik}
u_{n'}^{(k)+}(y)}{(\mu_n^{(i)}-im)(\mu^{(k)}_{n'} -im)}\label{A.8}\ee and for
$i\neq k$, Eq.(\ref{A.6}) reduces to (in matrix notations for  upper and lower
indices) \be \hat R= \hat \xi \hat R,\label{A.9}\ee where we have defined \be
\xi^{ik}_{nn'}= \frac{\int u_n^{(i)}(z)(-i\hat \partial- g \hat B-\mu_n^{(i)})
u^{(k)}_{n'} (z)
d^4z}{\mu_{n'}^{(k)}-im}\equiv\frac{V^{ik}_{nn'}}{\mu^{(k)}_{n'} -im},~~ i\neq
k \label{A.9a}\ee while by definition $ \xi^{ii}_{nn'}=0$. Introducing notation
 \be \eta^{(i)}_{nn'}\equiv \mu^{(i)}_n -im)
(\mu_{n'}^{(i)}-im) \int u_n^{(i)+}(x) S_0 (x,y) u_{n'}^{(i)} (y)
dxdy-(\mu_n^{(i)} -im) \delta_{nn'},\label{A.10}\ee for $R^{ii}$ one obtains
\be R^{ii}_{nn'} = \eta^i_{nn'}+ \xi_{nm}^{ij} R^{ji}_{mn'}.\label{A.11}\ee One
can rewrite (\ref{A.9}), (\ref{A.11}) as \be \hat R =\hat \eta +\xi \hat R, ~~
\hat R= \frac{1}{1-\hat \xi}\hat\eta.\label{A.12}\ee Denoting $\hat\xi \equiv
\hat V \frac{1}{\hat \mu-im}$, one can write finally $S$ in (\ref{A.7}) as \be
S(x,y)= S_0(x,y)+ \sum_{ik, nn' n''} u^{(i)}_n(x) \left(\frac{1}{\hat \mu- im -
\hat V}\right)_{nn'}^{ik} \left[\delta_{n'n''}+ (\mu_{n'}^{(k)} -im)
V^{ii}_{n'n''}\right] u_{n''}^{(k)}(y)\label{A.13}\ee (Note, that $(\hat
\mu)^{ik}_{nn'}= \mu^{(i)}_n \delta_{ik} \delta_{nn'}$).

Neglecting the second term in the square brackets in (\ref{A.13}) (which is
reasonable for zero and quasizero eigenvalues $\mu_n$), one can write \be
S(x,y) = S_0 (x,y) + \sum_{ik, nn'}u^{(i)}_n (x)\left(\frac{1}{\hat \mu- im -
\hat V}\right)_{nn'}^{ik} u_{n'}^{(k)+}(y)\label{A.14}\ee One can find the
eigenvalues $\Lambda_S$ of the operator $\hat \mu-im \hat V$, and
eigenfunctions $u_s(x)$, and as a result (\ref{A.14}) turn out to be \be S(x,y)
= S_0 (x,y) + \sum_s u_s (x) \frac{1}{\Lambda_s-im} u_s^+(y)\label{A.15}\ee and
$u_s(x)$ are collectivized eigenfunctions of the gas of top charges \be
\frac{1}{\hat \mu- im - \hat V} = U\frac{1}{\hat\Lambda- im } U^+;~~ u_s (x) =
U_{s,in} u_n^{(i)}(x)\label{A.16}\ee

For the zero net topcharge in the volume $V_4$ (with, say, periodic boundary
conditions and topcharges in the singular gauge) there are no global zero
modes, and only quasizero modes, which e.g. for dilute gas of instantons  of
size $\rho$, have  the Wigner semicircle  distribution  \cite{14,15} \be \nu
(\Lambda) = \frac{1}{\pi \bar V^2} (2N\bar V^2-\Lambda^2)^{1/2} \Theta(2N\bar
V^2- \Lambda^2),\label{A.17}\ee where the averaged value in the QCD instantonic
vacuum \cite{15}  $\bar V^2 \approx O\left(\frac{\rho^2}{V_4}\right),$~\\ ~$
\bar V^2 = \frac{2\kappa^2}{N},~~ \kappa\approx 0.14$ GeV.

In case with nonzero global topcharge $Q$, one obtains $N_Q$ zero modes with
$\Lambda_s=0$ in the sum (\ref{A.15}), and the sum (\ref{A.15}) is singular for
$m\to 0$. This fact is exploited in Section 4 to exemplify the new mechanism
for creation
small fermion masses.\\

\vspace{1cm}

 \setcounter{equation}{0}
\renewcommand{\theequation}{A2.\arabic{equation}}

{\bf \large
\noindent Appendix 2}\\

\noindent{\bf \large Diagonalization of the mass matrix and the CKM parametrization}\\

One can write the incident mass matrix obtained from the minima of wave
functional (\ref{18}) as

\be\hat V=\left(\begin{array}{lll} \mu_1&\mu_{12}&\mu_{13}\\
\mu_{12}&\mu_2&\mu_{23}\\
\mu_{13}&\mu_{23}&\mu_3\end{array}\right)\label{A2.1} \ee and we shall test our
assumption that nondiagonal elements $\mu_{ij}$ due to kink solutions are close
to "geometrical averages" of minima $\mu_i$ and $\mu_j$, (Coherent Mixing
Mechanism (CMM), namely \be \mu^2_{ij} = \mu_i \mu_j+ \Delta_{ij}=
\mu_i\mu_j(1+\eta_{ij}).\label{A2.2}\ee
The eigenvalue equation for (\ref{A2.1}) looks like \be (\mu_1-m) (\mu_2-m)
(\mu_3-m) + 2 \mu_{13}\mu_{12} \mu_{23}-\mu^2_{13} (\mu_2-m)
-\mu_{12}^2(\mu_3-m)-\mu_{23}^2(\mu_1-m)=0.\label{A2.3}\ee We assume that the
input masses $\mu_i$ satisfy either  condition I \be \mu_1<\mu_2<
\mu_3,\label{A2.4}\ee or more stringent condition II: \be \mu_1\ll \mu_2\ll
\mu_3.\label{A2.5}\ee Eq. (\ref{A2.3}) with the use of (\ref{A2.2}) can be
expanded  in powers of $ \eta_{ij}=\frac{\Delta_{ij}}{\mu_i\mu_j} $ \be m^3
-m^2 \sigma + m \xi - \zeta =0,\label{A2.6}\ee where $\sigma= \sum_i \mu_i,~~
 \xi=-\sum_{i\neq j}
\Delta_{ij},~~ \zeta = \mu_1\mu_2\mu_3(-\frac{1}{4} \sum_{i\neq j} \eta_{ij}^2
+\frac12 \sum_{ij, lk}\eta_{ij} \eta_{lk})$. Since we have connections between
roots $m_i$ and $\sigma,\xi,\zeta$, namely, \be m_1m_2m_3 = \zeta
;~~\sum_{i\neq j}m_im_j= \xi; ~\sum_i m_i =\sigma\label{A2.7}\ee one can
associate $\zeta$ with the smallest root $m_1$;  then putting $\zeta=0$, one
obtains from (\ref{A2.6}) $m_1 =0$ and for $m_2,m_3$ one has the equation \be
m^2 - m\sigma+ \xi =0\label{A2.8}\ee with the solutions:
$$ m=\frac{\sigma}{2} \pm \sqrt{\frac{\sigma^2}{4}-\xi}, ~~ m_2 =
\frac{\xi}{\sigma}+ \frac{\xi^2}{\sigma^3}
+O\left(\left(\frac{\xi^3}{\sigma^5}\right)^3\right),$$ \be
m_3=\sigma-\frac{\xi}{\sigma}
+O\left(\frac{\xi^2}{\sigma^3}\right).\label{A2.10}\ee To the first order in
$\zeta$  one obtains \be m_1= \frac{\zeta}{m_2m_3}, ~~ m_2
\cong\frac{\xi}{\sigma}, m_3 \cong \sigma \label{A2.10}\ee and conditions
$m_i>0$ yield $\xi>0,\zeta>0$. Using (\ref{A2.4}) or (\ref{A2.5}) one can
estimate \be m_1 \cong \frac{\zeta}{\xi},~~ m_2\approx \frac{\xi}{\mu_3}, ~~
m_3\approx \mu_3+\mu_2+\mu_1.\label{A2.11}\ee

Thus we conclude  that due to mixing the mass $m_3$ does not move significantly
from $\mu_3$, while the masses $m_1,  m_2$ can drastically decrease.

It is interesting to find out how the ratios of $m_i$ may change, when
$\Delta_{ik}$ or $\eta_{ik}$ are changing, i.e. we are interested in the motion
of the eigenvalues $m_i$ when mixing parameters $\Delta_{ik}, \eta_{ik}$ are
changing with $\mu_i$ fixed. To this end we keep two of $\eta_{ik}$ equal, e.g.
$\eta_{23} =\eta_{13} \equiv -\eta<0$, and $\eta_{12} \equiv -\delta,
|\delta|\ll \eta$. In this case we have $\zeta\cong \mu_1\mu_2\mu_3 \eta
\delta,~~ \xi= \eta \mu_3 (\mu_1+\mu_2)$, and \be m_1\cong \delta
\frac{\mu_1\mu_2}{\mu_1+\mu_2}\approx \delta \mu_1,~~ m_2 = \frac{\xi}{\sigma}
\cong \frac{\eta \mu_3 (\mu_1+\mu_2)}{\sigma}; \frac{m_1}{m_2} \approx
\frac{\delta}{\eta} \frac{\mu_1\mu_2}{(\mu_1+\mu_2)^2}\label{A2.12}\ee Thus one
can see that $\frac{m_1}{m_2}$ can be much smaller than $\frac{\mu_1}{\mu_2}$
for $\delta\ll \eta$, while $m_1$ can be made arbitrarily smaller than $\mu_1$,
and $m_2$ much smaller than $\mu_2$ for small enough $\delta$ and $\eta
(\delta\ll \eta)$.

Let us now turn to the unitary matrix $\widehat{W}$, which diagonalizes  the
mass matrix (\ref{A2.1}). At this point we can exploit the  results of recent
paper \cite{17}, where matrices $W_u, W_d$ are given for any form of the matrix
(\ref{A2.1}).

To make our analysis more general, in (\ref{A2.1}) we introduce also phases for
matrix elements $\mu_{ik}$: $ \arg \mu_{ik}=\delta_{ik}, ~~ ik =12, 13,23$ with
the condition $\delta_{12}-\delta_{13} + \delta_{23}=0$; then the matrices
which diagonalize the matrix $\mu_{ik}$ with the eigenvalues $m_1<m_2<m_3$ (so
that $\hat \mu= W^+ \hat m W$), can be readily written, using the general form
of the unitary matrix from \cite{17}.

\be
 W=\left( \begin{array}{lll}
 \frac{(\mu_2-m_1)
 (\mu_3-m_1)-|\mu_{23}|^2}{N_1},
&\frac{\mu_{13}\mu_{23}^*-\mu_{12}(\mu_3-m_2)}{N_2},&
 \frac{\mu_{12}\mu_{23}-\mu_{13}(\mu_2-m_3)}{N_3}\\
 &&\\
\frac{\mu^*_{13}\mu_{23} -\mu_{12}^* (\mu_3-m_1)}{N_1},&
 \frac{(\mu_1-m_2)(\mu_3-m_2) - |\mu_{13}|^2}{N_2} , &
 \frac{\mu_{12}^* \mu_{13}-\mu_{23} (\mu_1-m_3)}{N_3}\\
 &&\\

 \frac{\mu_{12}^* \mu_{23}^*-\mu_{13}^* (\mu_2-m_1)}{N_1} ,& \frac{\mu_{12}
 \mu_{13}^*-\mu_{23}^* (\mu_1-m_2)}{N_2}, & \frac{(\mu_1-m_3) (\mu_2-m_3)
 -|\mu_{12}|^2}{N_3}\end{array}\right)\label{A2.13a}\ee

Here we have used the notations similar to those from \cite{17}.

$$ N^2_1 = (m_3-m_1) (m_2-m_1) [ (\mu_2-m_1) (\mu_3 -m_1) - |\mu_{23}|^2]$$
\be  N^2_2 = (m_3-m_2) (m_2-m_1) [ (\mu_3-m_2) (m_2-\mu_1)
+|\mu_{13}|^2]\label{A2.14}\ee
$$N^2_3 = (m_3-m_2) (m_3-m_1) [ (m_3-\mu_1) (m_3-\mu_2)
-|\mu_{12}|^2].$$

For the case of CMM the masses $m_1 \approx \mu_1\delta,~~ m_2\approx
(\mu_1+\mu_2) \eta,$ $m_3\approx \mu_1+\mu_2+\mu_3,~~ 0<\delta\ll \eta <1$, and
one can estimate $W_{31}$ in the limit $\delta\to 0, ~~ (m_1\to 0$),
$W_{31}\approx -\frac12 \sqrt{\frac{\mu_3}{m_3}
}\sqrt{\frac{\mu_1}{\mu_1+\mu_2}} \sqrt{\frac{\mu_2}{\mu_3} }
\frac{\delta}{\eta}$. One can use this limit, $\delta\to 0 $, to simplify
considerably the matrix $W$, as it is done in \cite{17} if $W_{13}=0$. In our
case  $W$ has the same form as obtained in \cite{17} for $W^+$, so that  in our
notations we can write as\footnote{Note that the form (\ref{A2.13}) does not
contain the terms $O(\eta), O(\delta)$, which are present in (\ref{A2.13a}).}

\be W =\left( \begin{array}{lll} c_\alpha,& - s_\alpha c_\beta
e^{i\delta_{12}},& s_\alpha s_\beta e^{i\delta_{13}}\\
- s_\alpha
e^{-i\delta_{12}},& -c_\alpha c_\beta&, c_\alpha s_\beta e^{i\delta_{23}}\\
0, &s_\beta e^{-i\delta_{23}},& c_\beta\end{array}\right) \label{A2.13}\ee
where we have denoted \be c_\alpha=\sqrt{\frac{m_2-\mu_1}{m_2-m_1}},~~
s_\alpha=\sqrt{1-c^2_\alpha},
c_\beta=\sqrt{\frac{m_3-\mu_2}{2m_3-\mu_2-\mu_3}},~~
s_\beta=\sqrt{1-c^2_\beta}.\label{A2.16a}\ee

Note, that the matrix (\ref{A2.13}) contains 4 independent parameters:
$s_\alpha, s_\beta$ and two phases $\delta_{ik}$, as it should be for the
$3\times $ 3 unitary matrix. This form of $W, W^+$ is assumed both for $u$ and
$d$ sectors and the entries will be denoted as $c_\alpha^u, s_\alpha^u,
c_\beta^u, s_\beta^u$ and the same for $d$. As a result one can construct the
CKM matrix $V_{CKM} \equiv W_uW_d^+$.

Writing $V_{CKM} = \{V_{ik}\}$, one has for matrix elements.

$$V_{11} = c^u_\alpha c^d_\alpha+c^u_\beta c^d_\beta  s^u_\alpha s^d_\alpha
e^{i\Delta_{12}}+ s^u_\alpha s^u_\beta s^d_\alpha s^d_\beta e^{i\Delta_{13}},$$

$$ V_{12} = s_\alpha^uc^u_\alpha  c^d_\alpha c^d_\beta e^{i\delta_{12}^u}- c_\alpha^u
s^d_\alpha  e^{i\delta^d_{12}}+s^u_\alpha s^u_\beta c^d_\alpha s^d_\beta
e^{i\delta^u_{13}-i\delta^d_{23}}$$

$$V_{13}= s^u_\alpha s^u_\beta c^d_\beta e^{i\delta_{13}^u} - s^d_\beta s^u_\alpha c^u_\beta e^{i(\delta_{12}^u+\delta^d_{23})}$$

\be V_{21} =- s^u_\alpha  c^d_\alpha e^{-i\delta^u_{12}}+ c^u_\alpha c^u_\beta
c^d_\beta s^d_\alpha e^{-i\delta^d_{12}}+ c^u_\alpha s^u_\beta s^d_\alpha
s^d_\beta e^{i\delta^u_{23}-i\delta^d_{13}}\label{A2.17}\ee

$$V_{22}= s^u_\alpha   s^d_\alpha
 e^{-i\Delta_{12}}+ c_\alpha^u c_\beta^u c^d_\alpha
c^d_\beta + c^u_\alpha c^d_\alpha s_\beta^u s^d_\beta e^{i\Delta_{23}},$$

$$V_{23}=c^u_\alpha c^d_\beta s^u_\beta e^{i\delta^u_{23}}
- c^u_\alpha c^u_\beta  s^d_\beta  e^{i\delta^d_{23}},$$

$$V_{31}=c^u_\beta s^d_\alpha s^d_\beta  e^{-i\delta_{13}^d} - c^d_\beta s^u_\beta
s^d_\alpha e^{-i\delta^u_{23}-i\delta^d_{12}},$$

$$V_{32} = -  s^u_\beta c^d_\alpha
c^d_\beta e^{-i\delta_{23}^u} + c^d_\alpha c^u_\beta s^d_\beta
e^{-i\delta^d_{23}},$$

$$V_{33} =
s^u_\beta s^d_\beta e^{-i\Delta_{23}}+ c^u_\beta c^d_\beta.$$

Here notation is used $\Delta_{ik}\equiv\delta^u_{ik} -\delta^d_{ik}$.

One can also easily calculate the Jarlskog invariant $J$, \be J=Im (V_{23}
V_{33}^* V^*_{21} V_{31})\label{A2.18}\ee Using (\ref{A2.17}) one can write the
leading term $(c_\beta \approx c_\alpha = 1 , s_\beta\ll 1, s_\alpha\ll 1$) in
the form \be J= s^u_\alpha s^d_\alpha | s^d_\beta-s^u_\beta e^{i\Delta_{23}}|^2
\sin \Delta_{12}\label{A2.19}\ee

\vspace{1cm}

 \setcounter{equation}{0}
\renewcommand{\theequation}{A3.\arabic{equation}}

{\bf \large
\noindent Appendix 3}\\

\noindent{\bf \large Coherent mixing in four generations}\\

Here at first we simplify the problem, neglecting the phases $\delta_{ij}$ in
the $ 4\times 4$ mass matrix , taken in the form (\ref{45}), and define the
eigenvalues $m_i, i=1,2,3,4$ from the quartic equation \be m^4-a_1 m^3
+a_2m^2-a_3 m +a_4=0,\label{A3.1}\ee where the coefficients $a_i$ can be
expressed either via the eigenvalues: \be a_1 = \sum^4_{i=1} m_i, ~~a_2 =
\sum^4_{i\neq j=1} m_im_j,~~a_3 = \sum_{i\neq j\neq k} m_im_j m_k,~~a_4 =
\prod^4_{i=1} m_i,\label{A3.2}\ee or through the mass matrix coefficients:
$\mu_{ki}=\mu_{ik}=\sqrt{\mu_i\mu_k (1+\eta_{ik})}$. Expanding to the lowest
power of $\eta_{ik}$, one obtains \be a_1= \sum^4_{i=1} \mu_i,~~a_2 =
-\sum_{i\neq k}\mu_i\mu_k \eta_{ik}.\label{A3.3}\ee

\be a_3 =\prod^4_{i=1} \mu_i \sum^4_{n=1} \frac{1}{\mu_n}\sum_{i,k,l\neq n}
\left(\frac{1}{2} \eta_{il} \eta_{kl}
-\frac14\eta^2_{ik}\right).\label{A3.4}\ee

\be a_4=\frac14 \prod^4_{i=1} \mu_1\left\{\sum_{i\neq k\neq l} \eta_{ik}
\eta_{kl} \eta_{li} - \sum_{i\neq s \neq k\neq l} \eta_{ik} \eta_{kl} \eta_{ls}
+\sum_{i\neq j\neq l\neq k} \eta^2_{ij} \eta_{lk}\right\}=\label{A3.5}\ee
$$ =\frac14\prod^4_{i=1} \mu_1\left\{
\eta_{12} \eta_{13} \eta_{23}+ \eta_{14}( \eta_{23}^2- \eta_{13}\eta_{23}
-\eta_{12}\eta_{23})+\right.$$ $$ \eta_{24}( \eta_{13}^2- \eta_{12}\eta_{13}
-\eta_{13}\eta_{23})+ \eta_{34}( \eta_{12}^2- \eta_{12}\eta_{13}
-\eta_{12}\eta_{23})+$$
$$ \eta_{23}(
\eta_{14}^2- \eta_{14}\eta_{24} -\eta_{14}\eta_{34}+ \eta_{34}\eta_{24})
+\eta_{13}(\eta_{24}^2-
\eta_{24}\eta_{34}-\eta_{14}\eta_{24}+\eta_{34}\eta_{14})+$$ \be\left.
+\eta_{12}(\eta^2_{34} -\eta_{14} \eta_{34}-\eta_{24}\eta_{34}
+\eta_{24}\eta_{14})\right\}.\label{A3.6}\ee

In the last form  (\ref{A3.6}) the terms are ordered according to the power of
$\eta_{i4}$, which can tend to zero.

To simplify coefficients and establish a connection with the case of 3
generations, we assume that  only three new elements are nonzero in the
$4\times 4 ~\mu_{ik}$ as compared to $3\times 3 ~\mu_{ik}$, namely, $\mu_4,
\eta_{34} =\eta_{24} = -\bar \eta, \bar\eta>0$ in addition to considered before
$\eta_{13}=\eta_{23}=-\eta, ~~ \eta_{12} =-\delta$.

Then to the lowest order one has \be a_2
=\mu_1\mu_2\delta+\mu_3(\mu_1+\mu_2)\eta +\mu_4(\mu_2+\mu_3)\bar
\eta.\label{A3.7}\ee \be a_3=\mu_1\mu_2\mu_3\eta \delta -\frac14\mu_1\mu_2\mu_4
(\delta-\bar\eta)^2 -\frac14 \mu_1\mu_3\mu_4(\eta-\bar \eta)^2+\mu_2\mu_3\mu_4
\eta(\bar \eta-\frac{\eta}{4}),\label{A3.8}\ee \be a_4 =\frac34
\prod^4_{i=1}\mu_i \delta \eta\bar \eta.\label{A3.9}\ee

In   the case, when $\bar \eta>\frac{\eta}{4}$ and $\mu_4>\mu_3>\mu_2>\mu_1$,
all coefficients $a_i, ~~i=1,2,3,4$ are positive, yielding positive eigenvalues
$m_i$.  In the limit $\delta\to 0, \eta\to 0$ both  $a_3, a_4$ vanish and  for
two largest eigenvalues one has \be m_4, m_3= \frac{a_1}{2}\pm
\sqrt{\frac{a^2_1}{4}-a_2},\label{A3.10}\ee \be m_4 =\sigma
-\frac{\mu_3\mu_4\bar \eta}{\sigma}, ~~ \sigma
=\sum^4_{i=1}\mu_i,\label{A3.11}\ee \be m_3 = \frac{\mu_3\mu_4\bar
\eta}{\sigma}.\label{A3.12}\ee

Thus again, as in the case of three generations, the largest mass $m_4$ is
slightly larger than $\mu_4$, whereas $m_3\approx \bar \eta\mu_3$ can be
strongly shifted from the input value $\mu_3$. For $m_1, m_2$ one obtain an
estimate \be m_2\cong \frac{a_3}{m_3m_4} \cong \frac{a_3}{\bar \eta\mu_3\mu_4}
=\frac{\mu_2\eta (\bar \eta-\eta/4)}{\bar \eta}, ~~ m_1
=\frac{a_4}{m_2m_3m_4}\cong \frac{\frac34\mu_1 \delta\bar \eta}{\bar \eta
-\eta/4}.\label{A3.13}\ee

Now we have to find the unitary matrix $W$, which diagonalizes the $4\times 4$
matrix $\mu_{ik}$; for that we impose two conditions \be\hat W^+\hat W =\hat 1,
~~ W_{ki}^*W_{kl}=\delta_{il}\label{A3.14}\ee \be \hat W^+ \hat m \hat W=\hat
\mu,\label{A3.15}\ee where $\hat m=diag. (m_1, m_2, m_3, m_4)$.

In CMM the simplest $(4\times 4)$ form which is a natural extension of the
$(3\times 3)$ matrix (\ref{A3.15}) is \be W=\left(
\begin{array}{llll}
c_\alpha,& -s_\alpha c_\beta e^{i\delta_{12}}, & c_4 s_\alpha
s_{\beta} e^{i\delta_{13}},& s_4s_\beta s_\alpha e^{i\delta_{14}}\\
-s_\alpha e^{-i\delta_{12}},& -c_\alpha c_\beta,& c_4 s_\beta
c_\alpha e^{i\delta_{23}},& s_4 s_\beta c_\alpha e^{i\delta_{24}}\\
0,& s_\beta e^{-i\delta_{23}},& c_4 c_\beta, & s_4c_\beta
e^{i\delta_{34}}\\
0,&0,& -s_4e^{-i\delta_{34}},&c_4\\\end{array}\right).\label{A3.16}\ee The
phases $\delta_{ik}$ satisfy conditions \be
\delta_{24}=\delta_{23}+\delta_{34}, ~~ \delta_{14} =\delta_{13}+\delta_{34},
~~ \delta_{14}-\delta_{24}=\delta_{12}.\label{A3.17}\ee

One can notice that only two of these conditions are independent of the  old,
$3\times 3$ condition $\delta_{12}-\delta_{13}+\delta_{23}=0$. Here $s^2_4
+c^2_4=1$, so that in (\ref{A3.16}) one has two new parameters, e.g. $s_4$ and
$\delta_{34}$.

One can check, that the unitarity condition of $W$, Eq. (\ref{A3.14}), is
satisfied, while (\ref{A3.15}) yields approximately
$$ \mu_4\cong c^2_4m_4+ s^2_4c^2_\beta m_3+...$$
\be \mu_3\cong m_4s^2_4+ m_3 c^2_4c^2_\beta +...\label{A3.18}\ee
$$\mu_2\cong m_3 s^2_\beta +m_2,~~ \mu_1\cong
m_2s^2_\alpha+m_1c^2_\alpha.$$ From (\ref{A3.18}) one can see that $\mu_4<m_4$
and $\mu_i>m_i, i=1,2,3$ and the relations (\ref{A3.11}-\ref{A3.13}) are
approximately satisfied.

From (\ref{A3.15}) one can find $c_\alpha, c_\beta, c_4(s_i =\sqrt{1-c^2_i},
i=\alpha, \beta, 4)$:

\be c^2_\alpha =\frac{m_2-\mu_1}{m_2-m_1}, ~~ c^2_\beta
=\frac{m_3-\mu_2}{m_3+\mu_1- m_1-m_2},
~~c^2_4\cong\frac{\mu_4-m_3+\mu_2}{m_4-m_3+\mu_2}.\label{A3.19}\ee

\end{document}